\documentstyle[aps,prl,epsfig]{revtex}
\def\narrowtext{} \tighten
\twocolumn

\begin{document}
\draft
\title{ Spin Dynamics of the Magnetoresistive Pyrochlore Tl$_2$Mn$_2$O$_7$}
\author{J. W. Lynn and L. Vasiliu-Doloc}
\address{NIST Center for Neutron Research, National Institute of \\
Standards and Technology, Gaithersburg, Maryland 20899 and Center for\\
Superconductivity Research, Department of Physics, University of Maryland,\\
College Park, MD 20742}
\author{M. A. Subramanian}
\address{Dupont Central Research and Development, Experimental Station, Wilmington,\\
DE 19880}
\address{%
\begin{minipage}[t]{6.0in}
\begin{abstract}
Neutron scattering has been used to study the magnetic order and spin
dynamics of the colossal magnetoresistive pyrochlore Tl$_{2}$Mn$_{2}$O$_{7}$%
. On cooling from the paramagnetic state, magnetic correlations develop and
appear to diverge at $T_{C}$ ($123K$). In the ferromagnetic phase well
defined spin waves are observed, with a gapless ($\Delta <0.04$ meV)
dispersion relation $E=Dq^{2}$ as expected for an ideal isotropic
ferromagnet. As $T\rightarrow T_{C}$ from low $T$, the spin waves renormalize,
but no significant central diffusive component to the fluctuation spectrum
is observed in stark contrast to the La$_{1-x}$(Ca,Ba,Sr)$_{x}$MnO$_{3}$
system. These results argue strongly that the mechanism responsible for the
magnetoresistive effect has a different origin in these two classes of
materials.
\typeout{polish abstract}
\end{abstract}
\pacs{PACS numbers: 75.40.Gb, 75.70Pa, 75.30.Kz, 75.25.+z}
\end{minipage}}

\maketitle
\narrowtext

Pure LaMnO$_{3}$ is an antiferromagnetic insulator in which the Mn$^{3+}$O$%
_{3}$ octahedra exhibit a Jahn-Teller distortion that strongly couples the
magnetic and lattice system\cite{GeneralRef,Millis}. Doping with divalent
ions such as Ca, Sr, or Ba introduces Mn$^{4+}$, and with sufficient doping (%
$x\gtrsim 0.1$) the holes become mobile and the system transforms into a
metal. In this metallic regime the double-exchange mechanism allows
holes to move only if adjacent spins are parallel, which results in a dramatic
increase in the conductivity when the spins order ferromagnetically, either
by lowering the temperature or applying a magnetic field. The carrier
mobility is thus intimately tied to both the lattice and magnetism, and
considerable effort has been devoted to identifying the basic interactions
that dominate the energetics and control the magnetoresistive properties.
One avenue to unraveling these interactions is by measuring the spin
dynamics, and a number of anomalous features have been identified including
very strong damping of the spin waves in the ground state\cite{Doloc} and as
a function of temperature\cite{Perring}, anomalous spin wave dispersion\cite
{Hwang}, and the development of a strong spin-diffusion component to the
fluctuation spectrum well below $T_{C}$\cite{LynnPRL,Jaime,DolocMMM}.
Recently a new ``colossal'' magnetoresistive (CMR) compound has been
discovered, namely the pyrochlore Tl$_{2}$Mn$_{2}$O$_{7}$\cite
{Shimakawa,MasReview}, and an important question concerns whether this new
class of CMR materials contains the same underlying physics, or represents a
completely new and different CMR mechanism. We have investigated the
magnetic correlations, phase transition, and long wavelength spin dynamics
using neutron scattering techniques, and find no evidence of the anomalous
spin-diffusion component of the magnetic fluctuation spectrum that dominated
the phase transition in the optimally-doped La$_{1-x}$Ca$_{x}$MnO$_{3}$
(LCMO) manganites and appears to be associated with the spin component of
the polaron in these materials. These results, coupled with the absence of
any Jahn-Teller effects due to Mn$^{3+}$ in the system\cite
{Mas1,Shimakawa2,Kwei}, argues strongly that the Mn pyrochlore belongs to a
new class of CMR systems with a different underlying magnetoresistive
mechanism.

The neutron scattering measurements were carried out at the NIST Center for
Neutron Research. The polycrystalline sample weighed 1.5 g, and the
preparation technique, structure, and transport properties are described
elsewhere \cite{Mas1,Kwei}. The magnetic diffraction and inelastic data were
collected on the BT-9 triple axis spectrometer, with pyrolytic graphite
monochromator, analyzer, and filter, and a fixed energy of 13.7 meV. Higher
resolution inelastic data close to $T_{C}$ were taken on the SPINS spectrometer 
using a fixed final energy of 3.7 meV. Because
the long wavelength spin dynamics turns out to be approximately isotropic,
inelastic measurements on polycrystalline samples may be made in the forward
scattering direction (i.e. around the (000) reciprocal lattice point)
without loss in generality\cite{amorphous}. The inelastic measurements were
taken with horizontal collimations of 12$^{\prime }$-11$^{\prime }$-12$%
^{\prime }$-16$^{\prime }$ full width at half maximum (FWHM), while for the
diffraction data the collimations could be relaxed to improve the data rate.
The SANS data were collected on the NG-3 spectrometer using a wavelength of
5 \AA\ and a detector position of 8 m, with the intensity on the
two-dimensional position sensitive detector angularly averaged around the
beam-center position to obtain I($\left| {\bf q}\right| $). For the present
data the experimentally accessible $q$ range is 
$0.009\leq q\leq 0.12$ \AA $^{-1}$.

Tl$_{2}$Mn$_{2}$O$_{7}$ is cubic ($Fd\overline{3}m,$ $a$ = 9.892 \AA \  at room
tem-

\begin{center}
\begin{figure}[t]
\epsfig{file=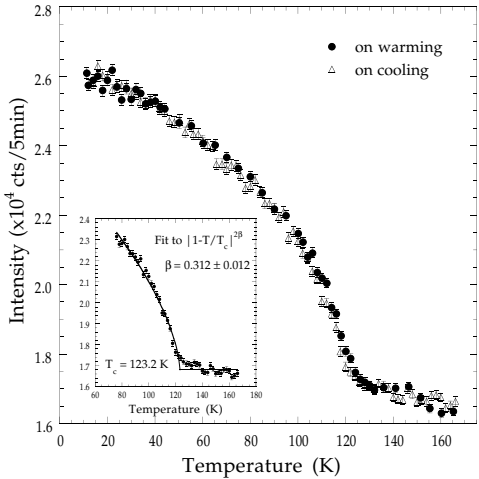,width=2.5in}
\vspace*{0.4cm}
\caption{Intensity of the ferromagnetic Bragg peak vs. temperature. No
significant difference is observed on warming and cooling. The inset shows a
power law fit, giving $T_{C}$=123.2 K and a critical exponent $\beta =0.312$.}
\end{figure}
\end{center}

\noindent perature), and the temperature dependence of the \{111\} Bragg peak is
shown in Fig. 1. Well above the ferromagnetic ordering temperature the
intensity originates from the structural Bragg peak, while the abrupt
increase signals the development of long range ferromagnetic order in the
sample. There is some additional scattering near $T_{C}$ due to critical
scattering, but for a powder sample this is quite a sharp transition
indicative of the high quality and uniformity of the sample. The inset shows
a fit of the data to a power law, which yields an estimate for the
transition temperature of $T_{C}=$ 123.2(3) K. The critical exponent we
obtain from the fit is $\beta =0.312(12)$, which is a typical value for a
three dimensional ferromagnet. The errors quoted are statistical only. The
ordered magnetic moment we observe in the ground state, based on the nuclear
and magnetic intensities of the first five Bragg peaks, is $\mu =2.907(36)$ $%
\mu _{B}$, which is very close to the $3$ $\mu _{B}$ expected for the Mn$%
^{4+}$ ion. This is a further indication \cite{Mas1,Ramirez} that there is
no significant concentration of Mn$^{3+}$ in the material, and hence no
possibility of the double-exchange mechanism being operative.

The development of magnetic order in this system can also be easily seen
in the SANS data, as shown in Fig. 2(a), where the intensity of the
scattering for $q$ = 0.012 \AA $^{-1}$ is shown as a function of temperature.
The data in the smallest $q$ regime exhibit an abrupt increase in the
scattering at $T_{C}$, and we interpret this as scattering due to domains
and domain walls. The intensity of this scattering follows the usual Porod
form ($I(q)\propto 1/Q^{4}$) indicating that the size of the objects that
are producing the scattering is larger than $2\pi /q\sim 600$ \AA. On the
other hand, in many ferromagnets this domain scattering is not observable in
this $q$ range, which indicates that the size of the domains in Tl$_{2}$Mn$%
_{2} $O$_{7}$ is relatively small.

At larger $q$ the scattering from the domains becomes weak and one can observe
the conventional critical scattering, as shown in Fig. 2(b). As the
temperature is cooled from above $T_{C}$, the scattering increases and peaks
near the ordering temperature. Below $T_{C}$ the scattering originates from
spin waves, and this scattering decreases with $T$ due to the decrease in the
number of spin waves. The scattering above $T_{C}$ is found to obey the
usual Lorentzian form $I(q)\propto 1/(q^{2}+\kappa ^{2})$, and the
correlation length $\xi =1/\kappa $ is plotted in Fig. 2(c). The range of
spin correlations increases as $T_{C}$ is approached, and appears to diverge
at the ordering temperature.

We now turn to the inelastic measurements of the spin fluctuation spectrum.
We find that the magnetic system behaves as an ideal isotropic ferromagnet
at low $T$, as is the case for all the metallic ferromagnetic manganites\cite
{Doloc,Perring,Hwang,LynnPRL,Jaime,Martin}. The magnetic excitations are
conventional spin waves, with a dispersion relation $E=\Delta +D(T)q^{2}$,
where $\Delta $ represents the spin wave energy gap and the spin stiffness
coefficient $D(T)$ is directly related to the exchange interactions. Fig. 3
shows a typical magnetic inelastic spectrum for a wave vector $q=0.13$ \AA
$^{-1}$. A flat background of 3 counts plus an elastic incoherent nuclear
peak of 61 counts, measured at 10 K, have been subtracted from these data. At
both temperatures shown we see well defined spin waves in energy gain (E%
\mbox{$<$}%
0) and energy loss (E%
\mbox{$>$}%
0). The solid curve is the result of a least-squares fit of the spin wave
cross section, convoluted with the instrumental resolution. At the higher
temperature the spin waves have renormalized to lower energy, broadened, and
the overall integrated intensity has increased, in agreement with
conventional spin wave theory. At each temperature we have taken data at a
series of $q$'s, and found that in this hydrodynamic (small $q$) regime the
dispersion obeys a quadratic law within experimental error, with a gap that
is too small to determine ($\Delta <0.04$ meV). Thus we conclude that Tl$%
_{2} $Mn$_{2}$O$_{7}$ behaves as a ``soft'' isotropic ferromagnet to an
excellent approximation.

The temperature dependence of the spin wave stiffness parameter $D(T)$ is
shown in Fig. 4. In the ground state we obtain a value of $D=39(1)$ meV$\cdot$\AA $%
^{2}$. The ratio of $D(0)/k_{B}T_{C}=$ 3.67 \AA $^{2}$ gives an estimate of
the range of the exchange interaction, and this result indicates that
nearest neighbor interactions dominate the energetics. At elevated
temperatures we find that the spin waves soften, and appear to collapse as $%
T\rightarrow T_{C}$; the data taken at 125 K exhibit spin diffusion, with a
spin-diffusion coefficient $\Lambda =11.5(8)$ meV$\cdot$\AA $^{2}$. This behavior
is in stark contrast to the behavior of the Ca-doped CMR materials, where a
quasielastic spin diffusion component develops as the Curie temperature is
approached, and in the range of optimal doping dominates the fluctuation
spectrum. In the LCMO case the spin waves do not appear to renormalize to
zero, and hence the ferromagnetic transition is not driven by the usual route of
the thermal 

\begin{center}
\begin{figure}[t]
\epsfig{file=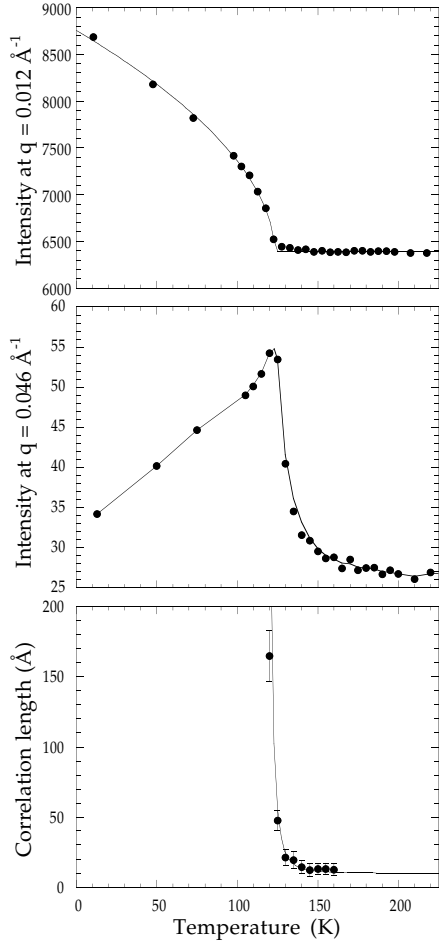,width=2.3in}
\vspace*{0.4cm}
\caption{Temperature dependence of the energy-integrated small angle magnetic
scattering (a) at small $q$, where the scattering is dominated by domain wall
scattering and thus follows a power of the magnetization (solid curve), (b)
at larger $q$, where the usual critical scattering in the vicinity of the
Curie temperature is observed. The solid curve is a guide only. (c) the
temperature dependence of the spin correlation length, showing that it rises
rapidly with decreasing temperature and appears to diverge at $T_{C}$. The
solid curve is a fit to a power law.}
\end{figure}
\end{center}

\noindent 
population of conventional spin waves. Rather, the
ferromagnetic phase transition appears to be driven by the quasielastic
component, which has been identified as the spin component of the polaron in
this system\cite{LynnMMM}. A central component to the fluctuation spectrum
has also been observed in the Sr doped\cite{DolocMMM} and Ba doped materials%
\cite{LynnBa}, as well as for Nd$_{0.7}$Sr$_{0.3}$MnO$_{3}$ and Pr$_{0.7}$Sr$%
_{0.3}$MnO$_{3}$\cite{Jaime}. In this latter work the authors suggested that
the strength of the central peak is inversely related to the value of $T_{C}$%
, and with the low value of $T_{C}$ of 123K for Tl$_{2}$Mn$_{2}$O$_{7}$ one
would then expect strong quasielastic scattering if the physics were the
same for these two classes of materials. The data in Fig. 3 clearly show
that we find{\bf \ }{\em no }evidence for a quasielastic component to the
fluctuation spectrum in the present material. There might still be a central
component to the fluctuation spectrum at smaller $q$ or for temperatures
closer to $T_{C}$ than have been investigated here, but it is clear that the
spin dynamics in the pyrochlore is fundamentally different than in the doped
manganites.

\begin{center}
\begin{figure}
\epsfig{file=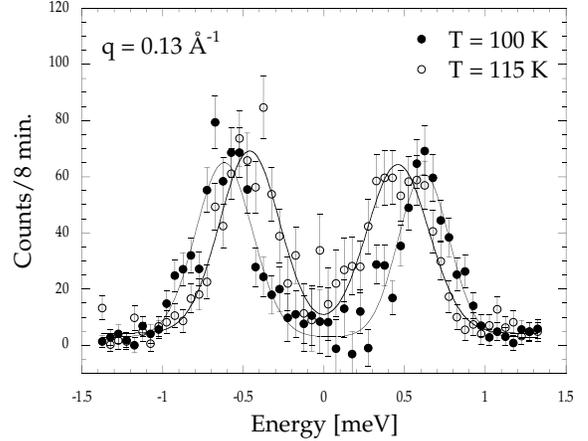,width=3.0in}
\vspace{0.1cm}
\caption{Inelastic spectrum at $q=0.13$ \AA $^{-1}$ for two temperatures below 
$T_{C}$. Spin waves are observed in energy gain ($E<0$) and energy loss ($E>0$%
). The solid curves are fits to a double Lorentzian spectral weight function
convoluted with the instrumental resolution. With increasing $T$ the spin
waves are seen to renormalize to lower energy, broaden, and increase in
intensity, as expected. No central component of the fluctuation spectrum is
observed, in strong contrast to the behavior of the La$_{1-x}$A$_{x}$MnO$%
_{3} $ systems.}
\end{figure}
\end{center}

The behavior for Tl$_{2}$Mn$_{2}$O$_{7}$ contrasts with that observed for
the doped LaMnO$_{3}$ manganite materials in the following important ways:
i) in the present system there is no evidence, from a variety of different
experimental techniques, of any significant Mn$^{3+}$ or associated
mixed-valent behavior, ii) there is also no significant structural
distortion that accompanies the development of a bulk magnetization, either
as a function of decreasing temperature or increasing applied magnetic
field, iii) the magnetic correlation length appears to grow and diverge at $T%
_{C}$ in the pyrochlore, while in the Ca-doped manganite the correlation
length is short ($\sim $12 \AA ) and only weakly temperature dependent\cite
{LynnPRL,SANSNature}, iv) for the spin dynamics, $D(T)$ appears to collapse
as the Curie temperature is approached from below, while for LCMO the spin
wave energies remain finite, and the phase transition appears to be
controlled by the development of a spin-diffusion central component to the
fluctuation spectrum. These results taken together argue convincingly that
the large magnetoresistance in the pyrochlore has a different origin than in
the manganites. In particular, band structure calculations\cite{Singh}
suggest that for Tl$_{2}$Mn$_{2}$O$_{7}$ the conductivity comes
predominantly from a Tl-O band rather than from the manganese lattice, and
hence the magnetic behavior and conduction arise from different sublattices.
However, the conduction bands turn out to be strongly spin differentiated in
the ferromagnetic state (as they are in the manganites) and this is thought
to produce the particularly strong spin scattering of the conduction
electrons that is responsible for the CMR effect. This separation of the
ferromagnetic lattice and the conduction band has been clearly revealed in
recent doping studies, where the resistivity and CMR was varied by
orders-of-magnitude with Sc substitution on the Tl site, while the magnetic
properties changed only modestly \cite{Ramirez}. With the very small
conduction electron density $n$ found in the pyrochlore\cite{Shimakawa,Mas1}%
, a model based on these observations has recently been developed, as a
low-density electron gas coupled to ferromagnetic spin fluctuations \cite
{Littlewood}. This model predicts a spin polaron regime, but only
sufficiently above $T_{C}$ (when $n\lesssim \kappa ^{3}$), and a
delocalization to an itinerant regime occurs due to the growth of
ferromagnetic correlations as $T_{C}$ is approached. The data of Fig. 2(c)
suggest that this crossover probably occurs quite close to $T_{C}$. Below $T%
_{C}$ significant spin polaron effects would not be expected in this theory,
in agreement with our observations. Direct observation of the proposed spin
polarons above $T_{C}$ will be quite difficult, though, because the strong
spin-diffusion scattering associated with the conventional paramagnetic
scattering will mask the small intensity associated with the low carrier
density. Single crystal samples will likely be necessary to explore this
possibility.

\begin{center}
\begin{figure}
\epsfig{file=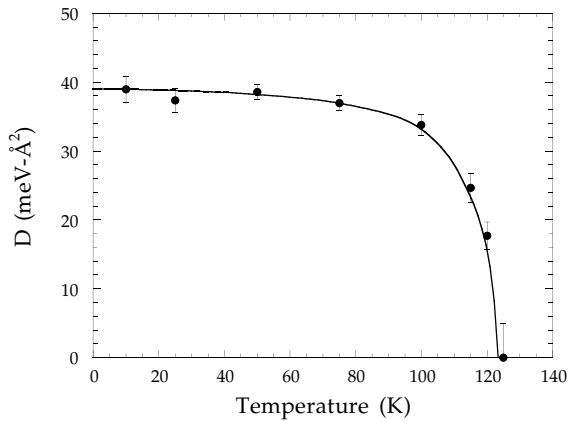,width=3.0in}
\vspace{0.1cm}
\caption{Temperature dependence of the spin wave stiffness $D(T)$, showing
that the spin waves renormalize in the usual way as the ferromagnetic
transition is approached. The solid curve is a guide only.}
\end{figure}
\end{center}

Research at the University of Maryland is supported by the NSF, DMR 97-01339
and NSF-MRSEC, DMR 96-32521. SPINS is supported in part by NSF, DMR 94-23101.


\begin{references}
\bibitem{GeneralRef}  For references to some of the earlier literature see,
for example, G. H. Jonker and J. H. Van Santen, Physica {\bf 16}, 337
(1950); {\em ibid }{\bf 19, }120 (1950); E. O. Wollan and W. C. Koehler,
Phys. Rev. {\bf 100}, 545 (1955); C. Zener, Phys. Rev. {\bf 81}, 440 (1951), 
{\em ibid }{\bf 82}, 403 (1951); J. B. Goodenough, Phys. Rev. {\bf 100}, 564
(1955); P. W. Anderson and H. Hasegawa, Phys. Rev. {\bf 100}, 675 (1955).

\bibitem{Millis}  A.J. Millis, P.B. Littlewood, and B.I. Shraiman, Phys.
Rev. Lett. {\bf 74}, 5144 (1995); A.J. Millis, Phys. Rev. B {\bf 55}, 6405
(1997).

\bibitem{Doloc}  L. Vasiliu-Doloc, J. W. Lynn, A. H. Moudden, A. M. de
Leon-Guevara, and A. Revcolevschi (preprint)

\bibitem{Perring}  T. G. Perring, G. Aeppli, S. M. Hayden, S. A. Carter, J.
P. Remeika, and S-W. Cheong, Phys. Rev. Lett. {\bf 77}, 711 (1996).

\bibitem{Hwang}  H. Y. Hwang, P. Dai, S-W. Cheong, G. Aeppli, D. A. Tennant,
and H. A. Mook, Phys. Rev. Lett. {\bf 80}, 1316 (1998).

\bibitem{LynnPRL}  J. W. Lynn, R. W. Erwin, J. A. Borchers, Q. Huang, A.
Santoro, J. L. Peng, and Z. Y. Li, Phys. Rev. Lett. {\bf 76}, 4046 (1996).

\bibitem{Jaime}  J. A. Fernandez-Baca, P. Dai, H. Y. Hwang, S-W. Cheong, and
C. Kloc, Phys. Rev. Lett. {\bf 80}, 4012 (1998).

\bibitem{DolocMMM}  L. Vasiliu-Doloc, J. W. Lynn, Y. M. Mukovskii, A. A.
Arsenov, and D. A. Shulyatev, J. Appl. Phys. {\bf 83} (1998) (in press).

\bibitem{Shimakawa}  Y. Shimakawa, Y. Kubo, and T. Manako, Nature {\bf 379},
53 (1996).

\bibitem{MasReview}  For a review of the pyrochlores, see M. A. Subramanian
and A. W. Sleight, Chapter 107 in {\em Handbook on the Physics and Chemistry
of Rare Earths}, Vol. 16, Ed. by K. A. Geschneider, Jr. and L. Eyring
(Elsevier, New York 1993).

\bibitem{Mas1}  M. A. Subramanian, B. H. Toby, A. P. Ramirez, W. J.
Marshall, A. W. Sleight, and G. H. Kwei, Science {\bf 273}, 81 (1996).

\bibitem{Shimakawa2}  Y. Shimakawa, Y. Kubo, T. Manako, Y. V. Sushko, D. N.
Argyriou, and J. D. Jorgensen, Phys. Rev. B {\bf 55}, 6399 (1997)

\bibitem{Kwei}  G. H. Kwei, C. H. Booth, F. Bridges, and M. A. Subramanian,
Phys. Rev. B {\bf 55}, R688 (1997).

\bibitem{amorphous}  For a recent review of the experimental technique see
J. W. Lynn and J. A. Fernandez-Baca, Chapter 5 in {\em The Magnetism of
Amorphous Metals and Alloys}, ed. by J. A. Fernandez-Baca and W-Y. Ching
(World Scientific, New Jersey (1995), p. 221.

\bibitem{Ramirez}  A. P. Ramirez and M. A. Subramanian, Science {\bf 277},
546 (1997).

\bibitem{Martin}  M. C. Martin, G. Shirane, Y. Endoh, K. Hirota, Y.
Moritomo, and Y. Tokura, Phys. Rev. B {\bf 53}, 14285 (1996).

\bibitem{LynnMMM}  J. W. Lynn, R. W. Erwin, J. A. Borchers, Q. Huang, A.
Santoro, J. L. Peng, and R. L. Greene, J. Appl. Phys. {\bf 81}, 5488 (1997).

\bibitem{LynnBa}  J. W. Lynn, L. Vasiliu-Doloc, S. Skanthakumar, S. N.
Barilo, G. L. Bychkov, and L. A. Kurnevitch (preprint).

\bibitem{SANSNature}  J. M. De Teresa, M. R. Ibarra, P. A. Algarabel, C.
Ritter, C. Marquina, J. Blasco, J. Garcia, A. del Moral, and Z.Arnold,
Nature {\bf 386}, 256 (1997).

\bibitem{Singh}  D. J. Singh, Phys. Rev. B {\bf 55}, 313 (1997).

\bibitem{Littlewood}  P. Majumdar and P. Littlewood (preprint).

\newpage
\end{references}
\end{document}